\documentclass[prl,aps,superscriptaddress,twocolumn,notitlepage,showpacs,nofootinbib]{revtex4-1}
\usepackage{latexsym}
\usepackage{amsmath}
\usepackage{amssymb}
\usepackage{graphicx}
\usepackage{caption}
\usepackage{subfigure}
\usepackage{float}
\usepackage{mathrsfs}
\usepackage{color}
\usepackage{txfonts}
\usepackage[justification=centering,
            format=plain]{caption}

\renewcommand{\raggedright}{\leftskip=0pt \rightskip=0pt plus 0cm}

\begin{document}

\title{Quantum drives produce strong entanglement between YIG samples without using intrinsic nonlinearities}

\author{Jayakrishnan M. Prabhakarapada Nair}
\email{jayakrishnan00213@tamu.edu}
\affiliation{Institute for Quantum Science and Engineering, Texas A$\&$M University, College Station, TX 77843, USA}

\author{G. S. Agarwal}
\email{girish.agarwal@tamu.edu}
\affiliation{Institute for Quantum Science and Engineering, Texas A$\&$M University, College Station, TX 77843, USA}
\affiliation{Department of Biological and Agricultural Engineering, Department of Physics and Astronomy, Texas A$\&$M University, College Station, TX 77843, USA}

\date{\today}

\begin{abstract}
We show how to generate an entangled pair of yttrium iron garnet (YIG) samples in a cavity-magnon system without using any nonlinearities which are typically very weak. This is against the conventional wisdom which necessarily requires strong Kerr like nonlinearity. Our key idea, which leads to entanglement, is to drive the cavity by a weak squeezed vacuum field generated by a flux-driven Josephson parametric amplifier (JPA). The two YIG samples interact via the cavity. For modest values of the squeezing of the pump, we obtain significant entanglement. This is the principal feature of our scheme. We discuss entanglement between macroscopic spheres using several different quantitative criteria. We show the optimal parameter regimes for obtaining entanglement which is robust against temperature. We also discuss squeezing of the collective magnon variables.
\end{abstract}

\maketitle



Yttrium iron garnet (YIG), an excellent ferrimagnetic system, has attracted considerable attention during the past few years. The Kittel mode \cite{1} in YIG possesses unique properties including rich magnonic nonlinearities \cite{2} and a low damping rate \cite{3} and in addition the high spin density in YIG allows strong coupling between magnons and microwave cavity photons giving rise to quasiparticles, namely the cavity-magnon polaritons \cite{3,4,5,6,7,8}. Strong coupling between the YIG sphere and the cavity photons have been observed at both cryogenic and room temperatures \cite{8}. Aided by these superior properties, YIG is reckoned to be the key ingredient in future quantum information networks \cite{rev}. Thus a variety of intriguing phenomena have been investigated in the context of magnons. This include the observation of bistability \cite{9}, cavity spintronics \cite{7,n2}, level attraction for cavity magnon-polaritons \cite{n3}, magnon dark modes \cite{10}, the exceptional point \cite{11} etc. By virtue of the strong coupling among magnons, a multitude of quantum information aspects have been investigated including the coupling of magnons to a superconducting qubit \cite{12} and phonons \cite{13}. Other interesting phenomena involve magnon induced transparency \cite{14}, magnetically controllable slow light \cite{15} etc. 

Owing to the diverse interactions of magnons with other information carriers, YIG offer a novel platform in the analysis of macroscopic quantum phenomena. The coherent phonon-magnon interactions due to the radiation pressure like magnetostrictive deformation \cite{16} was studied. The nonlinear interaction between magnons and phonons can give rise to magnomechanical entanglement which further transfers to photon-magnon and photon-phonon subsystems, generating a tripartite entangled state \cite{17}. Another recent work proposed a scheme to create squeezed states of both magnons and phonons in a hybrid magnon-photon-phonon system \cite{18}. The squeezing generated in the cavity was transferred to the magnons via the cavity-magnon beamsplitter interaction. 

There is not much work on the coupling of two macroscopic YIG samples in a cavity. Recently the spin current generation in a YIG sample due to excitation in another YIG sample has been investigated \cite{n2}. This arises from the cavity mediated coupling between the two samples.  It is thus natural to consider the possibility of quantum entanglement between two YIG samples as there has been significant interest in the study of quantum entanglement between macroscopic systems. Recently there has been remarkable success in the observation of quantum entanglement between macroscopic mechanical oscillators \cite{22,23} with photonic crystal cavities and with superconducting quibits.  In addition entanglement between cavity field and mechanical motion has been reported \cite{21}. The conventional wisdom of producing entanglement involves nonlinearities in the system. The well known nonlinearities are the magnetostrictive interaction \cite{13} and the Kerr effect \cite{2}. The magnetostrictive force allows the magnons to couple to the phonons and can be used to generate magnon-phonon entanglement \cite{17}. The Kerr nonlinearity arises from the magnetocrystalline anisotropy and has been used to produce bistability in magnon-photon systems. In recent publications, these nonlinearities have been used to produce entanglement between two magnon modes in a magnon-cavity system \cite{n4,n5,foot1}. 

 Here we present a scheme to generate an entangled pair of YIG spheres in a cavity-magnon system without using any nonlinearities. In addition, we also investigate the squeezed states of the coupled system of two YIG spheres. Two YIG spheres are coupled to the cavity field and the cavity is driven by a squeezed vacuum field \cite{24,25}, resulting in a squeezed cavity field. A flux-driven Josephson parametric amplifier (JPA) is used to generate the squeezed vacuum microwave field. The squeezing in the cavity will be transferred to the two YIG samples due to the cavity-magnon beamsplitter interaction. Based on experimentally attainable parameters, we show that significant bipartite entanglement can be generated between the YIG samples. The entanglement is robust against temperature. Our results can be extended to other geometries of YIG. Further the method that we propose is quite generic and can be used for other macroscopic systems. 
  
\begin{figure}[!t]
 \captionsetup{justification=raggedright,singlelinecheck=false}
\centering
\includegraphics[scale=0.1]{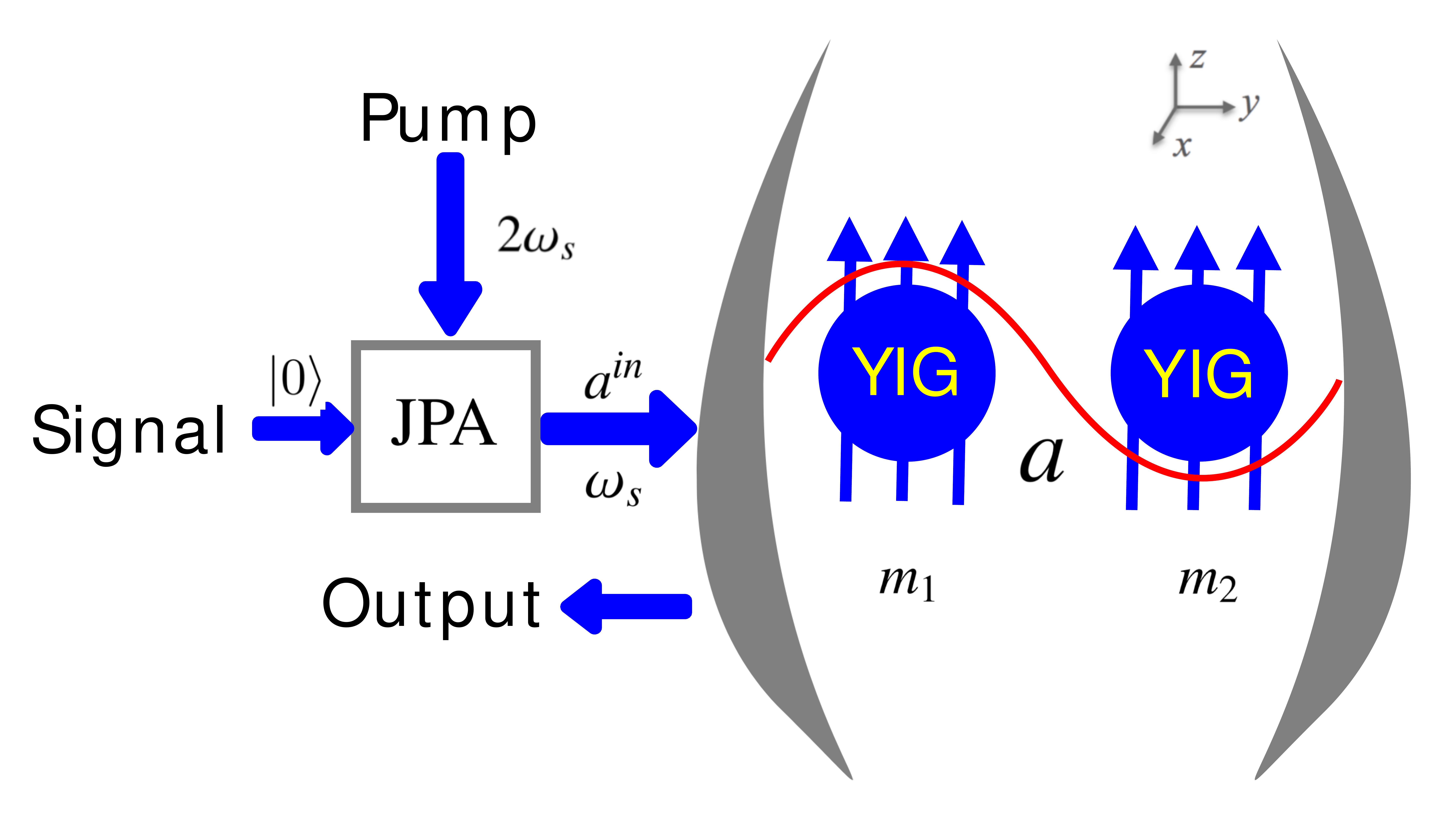}
\caption{\small  {Two YIG spheres are placed inside a microwave cavity near the maximum magnetic field of the cavity mode, and simultaneously in a uniform bias magnetic field. The cavity is driven by a week squeezed vacuum field generated by a flux-driven JPA. The magnetic field of the cavity mode is in the $\textit{x}$ direction and the bias magnetic field is applied along the $\textit{z}$ direction.}} \label{fna} 
\end{figure}

We consider the cavity-magnon system \cite{13,16,17} which consists of cavity microwave photons and magnons, as shown in figure \ref{fna}. The magnons are quasiparticles,  a collective excitation of a large number of spins in a YIG sphere. They are coupled to the cavity photons via the magnetic dipole interaction. The Hamiltonian of the system reads \cite{foot}
\begin{align}\
\mathcal{H}/\hbar&=\omega_{a} a^{\dagger}a+\omega_{m_{1}}m_{1}^{\dagger}m_{1} + \omega_{m_{2}}m_{2}^{\dagger}m_{2} \notag\\ 
&+g_{m_{1}a}(a+a^{\dagger})(m_{1}+m_{1}^{\dagger})+g_{m_{2}a}(a+a^{\dagger})(m_{2}+m_{2}^{\dagger}), \label{1}
\end{align}
where $a$ $(a^{\dagger})$ are the annihilation (creation) operator of cavity mode, $m_{1}$, $m_{2}$ ( $m_{1}^{\dagger}$, $m_{2}^{\dagger}$) are the annihilation (creation) operators of the two magnon modes and they represent the collective motion of spins via the Holstein-Primakoff transformation \cite{26} in terms of Bosons. The parameters $\omega_{a}$, $\omega_{m_{i}}$ (\textit{i}=1,2) are the resonance frequencies of the cavity and the magnon modes. Hereafter, wherever we use a subscript `$i$' it can take values from $1$ to $2$. The magnon frequency is given by the expression $\omega_{m_{i}}=\gamma H_{i}$, where $\gamma/2\pi=28$ GHz/T is the gyromagnetic ratio and $H_{i}$ are the external bias magnetic fields. The $g_{m_{i}a}$ in Eq.(\ref{1}) are the linear photon-magnon coupling strengths. The cavity is driven by a week squeezed vacuum field generated by a flux driven JPA. JPAs can in principle amplify a single signal quadrature without adding any extra noise. The squeezed vacuum is generated by degenerate parametric down-conversion using the nonlinear inductance of Josephson junctions \cite{28,29,30,31,32,33,34,35,36,37,38} and a squeezing down to 10\% of the vacuum variance has been produced \cite{30}. The operation of generating squeezed vacuum is depicted in figure \ref{fna}. Vacuum fluctuations are at the signal port and the pump field is applied at frequency 2$\omega_{s}$. The pump photon splits into a signal and an idler photon. Strong quantum correlations between the signal and idler photons are generated which result in squeezing. The output is at the frequency $\omega_{s}$ \cite{31,34}. The Hamiltonian described by Eq.(\ref{1}) does not contain terms involving the input drive field. We use standard quantum Langevin formalism to model the system and the equations describing the evolution of the system operators will contain the input drive terms. Applying the rotating-wave approximation $g_{m_{i}a}(a+a^{\dagger})(m_{i}+m_{i}^{\dagger})$ becomes $g_{m_{i}a}(am^{\dagger}+a^{\dagger}m)$ \cite{3,4,5,6,7,13}. In the rotating frame at the frequency $\omega_{s}$ of the squeezed vacuum field, the quantum Langevin equations (QLEs) describing the system can be written as follows
\begin{align}
 \dot{a}&=-(i\Delta_{a}+k_{a})a-ig_{m_{1}a}m_{1}-ig_{m_{2}a}m_{2}+\sqrt{2k_{a}}a^{in},\notag\\
 \dot{m_{1}}&=-(i\Delta_{m_{1}}+k_{m_{1}})m_{1}-ig_{m_{1}a}a+\sqrt{2k_{m_{1}}}m_{1}^{in}, \label{2} \\
 \dot{m_{2}}&=-(i\Delta_{m_{2}}+k_{m_{2}})m_{2}-ig_{m_{2}a}a+\sqrt{2k_{m_{2}}}m_{2}^{in},\notag
\end{align}
where $\Delta_{a}=\omega_{a}-\omega_{s}$, $\Delta_{m_{i}}=\omega_{m_{i}}-\omega_{s}$, $k_{a}$ is the dissipation rate of the cavity, $k_{m_{i}}$ are the dissipation rates of the magnon modes, and $a^{in}$, $m_{i}^{in}$ are the input noise operators of the cavity and magnon modes, respectively. The input noise operators are characterized by zero mean and the following correlation relations \cite{27}, $\langle a^{in}(t)a^{in\dagger}(t^\prime)\rangle=(\cal{N}+$1)$\delta (t-t^\prime)$, $\langle a^{in\dagger}(t)a^{in}(t^\prime)\rangle=\cal{N}$$\delta (t-t^\prime)$, $\langle a^{in}(t)a^{in}(t^\prime)\rangle=\cal{M}$$\delta (t-t^\prime)$,  $\langle a^{in\dagger}(t)a^{in\dagger}(t^\prime)\rangle=\cal{M^{*}}$$\delta (t-t^\prime)$, where $\cal{N}$=$\sinh^2 {r}$, $\cal{M}$=$e^{i\theta}\sinh{r}\cosh{r}$ with $r$ and $\theta$ being the squeezing parameter and the phase of the input squeezed vacuum field, respectively.  We have the other input correlations for the magnon as $\langle m_{i}^{in}(t)m_{i}^{in\dagger}(t^\prime)\rangle=[N_{m_{i}}(\omega_{m_{i}})+1]\delta (t-t^{\prime})$, $\langle m_{i}^{in\dagger}(t)m_{i}^{in}(t^\prime)\rangle =N_{m_{i}}(\omega_{m_{i}})\delta (t-t^{\prime})$, where $N_{m_{i}}(\omega_{m_{i}})=[\text{exp}(\frac{\hbar \omega_{m_{i}}}{k_{B}T})-1]^{-1}$ are the equlibrium mean thermal magnon numbers of the two magnon modes.

We now show that the YIG spheres can be entangled by resonantly driving the cavity with a squeezed vacuum field. We write down the field operators as their steady state values plus the fluctuations around the steady state. The fluctuations of the system can be described by the QLEs
\begin{align}
\delta \dot{a}&=-(i\Delta_{a}+k_{a})\delta a-ig_{m_{1}a}\delta m_{1}-ig_{m_{2}a}\delta m_{2}+\sqrt{2k_{a}}a^{in},\notag\\
\delta \dot{m_{1}}&=-(i\Delta_{m_{1}}+k_{m_{1}})\delta m_{1}-ig_{m_{1}a}\delta a+\sqrt{2k_{m_{1}}}m_{1}^{in}, \label{3} \\
\delta \dot{m_{2}}&=-(i\Delta_{m_{2}}+k_{m_{2}})\delta m_{2}-ig_{m_{2}a}\delta a+\sqrt{2k_{m_{2}}}m_{2}^{in}.\notag
\end{align}
The quadratures of the cavity field and the two magnon modes are given by $\delta X=(\delta a+\delta a^{\dagger})/ \sqrt{2}$, $\delta Y=i(\delta a^{\dagger}-\delta a)/\sqrt{2}$, $\delta x_{i}=(\delta m_{i}+\delta m_{i}^{\dagger})/\sqrt{2}$ and $\delta y_{i}=i(\delta m_{i}^{\dagger}-\delta m_{i})\sqrt{2}$, and similarly for the input noise operators. The QLEs describing the quadrature fluctuations $(\delta X, \delta Y, \delta x_{1}, \delta y_{1}, \delta x_{2}, \delta y_{2})$ can be written as 
\begin{align}
\dot{u}(t)=Au(t)+n(t), \label{4}
\end{align}
where $u(t)=[\delta X(t),\delta Y(t), \delta x_{1}(t),\delta y_{1}(t), \delta x_{2}(t), \delta y_{2}(t)]^{T}$, $n(t)=[\sqrt{2k_{a}}X^{in}, \sqrt{2k_{a}}Y^{in}, \sqrt{2k_{m_{1}}}x_{1}^{in},\sqrt{2k_{m_{1}}}y_{1}^{in}, \sqrt{2k_{m_{2}}}x_{2}^{in}, \\\sqrt{2k_{m_{2}}}y_{2}^{in} ]^{T}$ and
 \begin{align}
 A=
  \begin{bmatrix}
    -k_{a} & \Delta_{a} & 0 & g_{m_{1}a} & 0 & g_{m_{2}a} \\
    -\Delta_{a} & -k_{a} & -g_{m_{1}a} & 0 & -g_{m_{2}a} & 0\\
    0 & g_{m_{1}a} & -k_{m_{1}} & \Delta_{m_{1}} & 0 & 0\\
    -g_{m_{1}a} & 0 & -\Delta_{m_{1}} & -k_{m_{1}} & 0 & 0\\
    0 & g_{m_{2}a} & 0 & 0 & -k_{m_{2}} & \Delta_{m_{2}}\\
    -g_{m_{2}a} & 0 & 0 & 0 & -\Delta_{m_{2}} & -k_{m_{2}}
  \end{bmatrix}.
  \label{6}
\end{align}
The system is a continuous variable (CV) three- mode Gaussian state and it can be completely described by a $6\times6$ covariance matrix (CM) $V$ defined as $V$$(t)$$=\frac{1}{2}\langle u_{i}(t)u_{j}(t^{\prime})+u_{j}(t^{\prime})u_{i}(t)\rangle$, (\textit{i}, \textit{j} =1, 2....6). The steady state CM $V$ can be obtained by solving the Lyapunov equation \cite{39,40}
\begin{figure}[!t]
\captionsetup{justification=raggedright,singlelinecheck=false}
\centering
\includegraphics[scale=0.23]{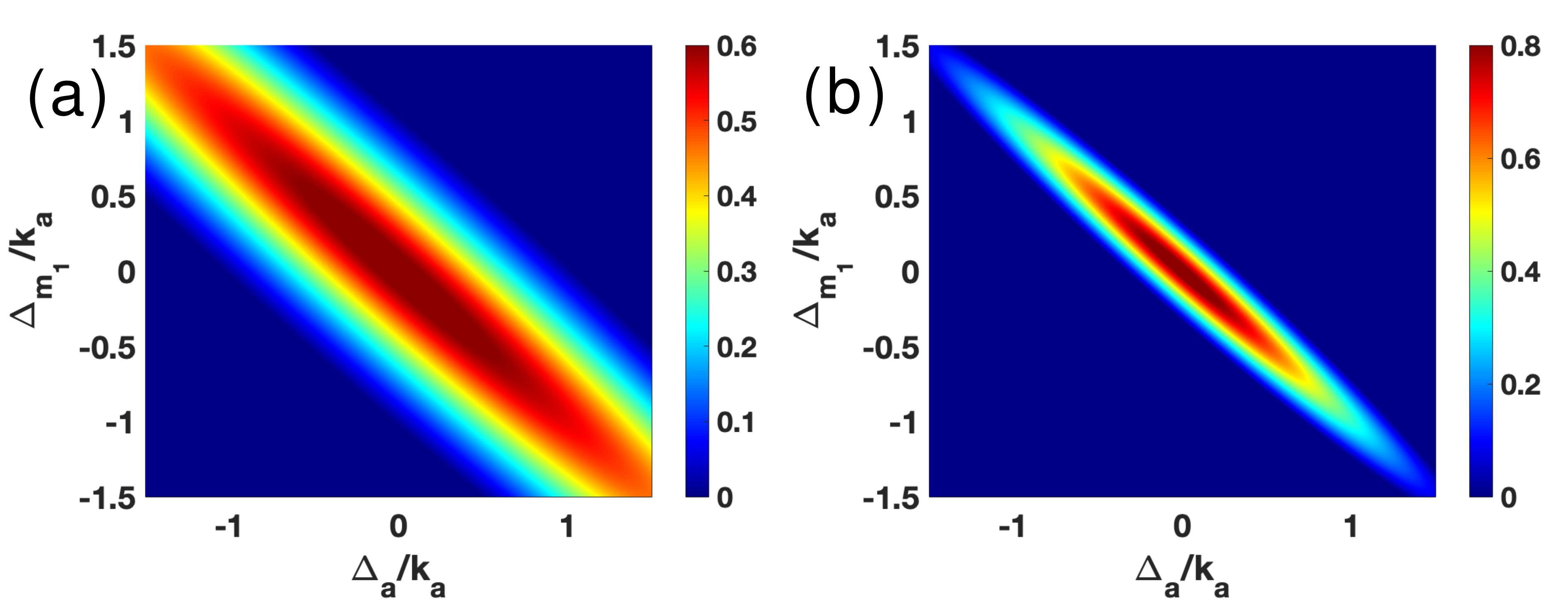}
\caption{\small  {Density plot of bipartite entanglement $E_{m_{1}m_{2}}$ between the two magnon modes versus $\Delta_{a}$ and $\Delta_{m_{1}}$ (a) with $\Delta_{m_{2}}=\Delta_{m_{1}}$, $r=1$, $\theta=0$, $T=20$ mK, (b) with $\Delta_{m_{2}}=\Delta_{m_{1}}$, $r=2$, $\theta=0$, $T=20$ mK. Other parameters are given in the text.}}\label{f2a} 
\includegraphics[width=8.6 cm, height=6 cm]{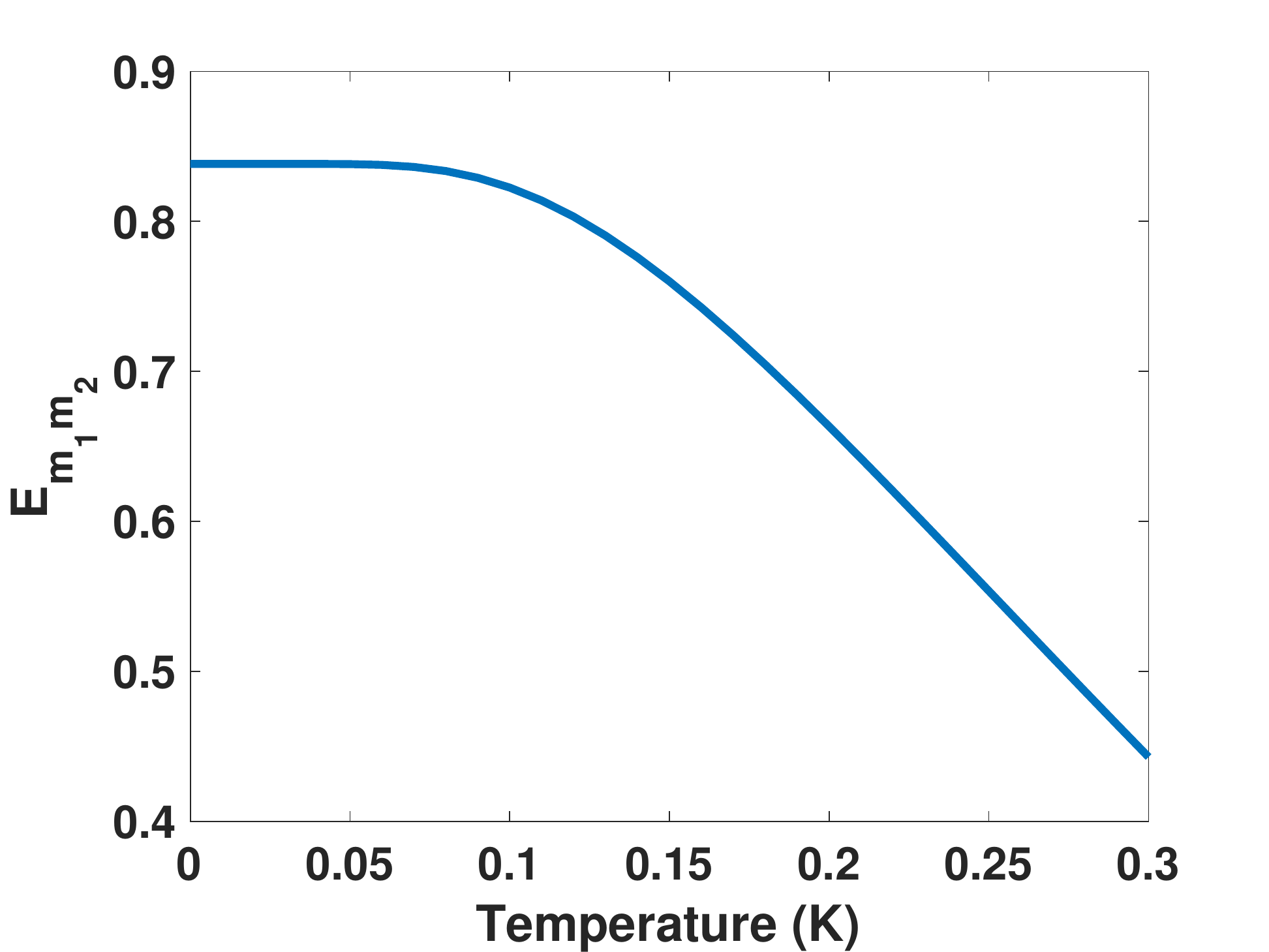}
\caption{\small  {Plot of bipartite entanglement $E_{m_{1}m_{2}}$ between the two magnon modes against temperature with $\Delta_{a}=\Delta_{m_{1}}=\Delta_{m_{2}}=0$, $r=2$ and $\theta=0.$}}\label{f2b}
\end{figure}
\begin{align}
AV+VA^{T}=-D, \label{7}
\end{align}

where $D$ is the diffusion matrix defined as $\langle n_{i}(t)n_{j}(t^{\prime})+n_{j}(t^\prime)n_{i}(t)\rangle /2=D_{ij}\delta (t-t^{\prime})$. We use logarithmic negativity \cite{n1} as the quantitative measure to investigate the bipartite entanglement $E_{m_{1}m_{2}}$ between the two magnon modes.  It can be obtained from $E_{m_{1}m_{2}}=max[0,-ln(2\nu_{-})]$ where $\nu_{-}=min[eig(i\Omega P_{12}VP_{12})]$, $\Omega=i\sigma_{y}\bigoplus i\sigma_{y}$, $P_{12}=1\bigoplus \sigma_{z}$ and $\sigma_{y}$, $\sigma_{z}$ are the Pauli matrices \cite{43}. Figure \ref{f2a}(a)-(b) shows the bipartite entanglement between the two magnon modes at two different squeezing parameters. We use a set of experimentally feasible parameters \cite{13}: $\omega_{a}/2\pi=10$ GHz, $k_{a}/2\pi=5k_{m_{i}}/2\pi=5$ MHz, $g_{m_{1}a}=g_{m_{2}a}=4k_{a}$ and $T=20$ mK, $N_{m_{1}}=N_{m_{2}}\approx0$ at 20 mK. The YIG sphere has a diameter  $250$-$\mu m$ and the number of spins $N\approx3.5\times 10^{16}$. We have adopted the parameters so that the two magnon modes are identical. We observe that $\Delta_{a}=\Delta_{m_{1}}=\Delta_{m_{2}}=0$, in other words $\omega_a=\omega_s$, $\omega_{m_{i}}=\omega_s$ are optimal for the entanglement between the two YIG samples.
At resonance we observe the maximum amount of entanglement and it increases with the increase in the squeezing parameter. Figure \ref{f2b} shows that the bipartite entanglement is quite robust against temperature. We observe significant amount of entanglement even at $T=$0.5 K which is quite remarkable for the system of two YIG spheres. We have chosen identical coupling between photon and the two magnon modes. In the case of unequal coupling the entanglement goes down. Although we have chosen two identical YIG spheres, one can have two cuboidal YIG samples as in \cite{n2} with an angle $\theta$ between the external magnetic field and the local microwave magnetic field at one YIG sample. This makes the resonance frequencies of the two samples different. 
\begin{figure}[!t]
 \captionsetup{justification=raggedright,singlelinecheck=false}
\centering
\includegraphics[scale=0.23]{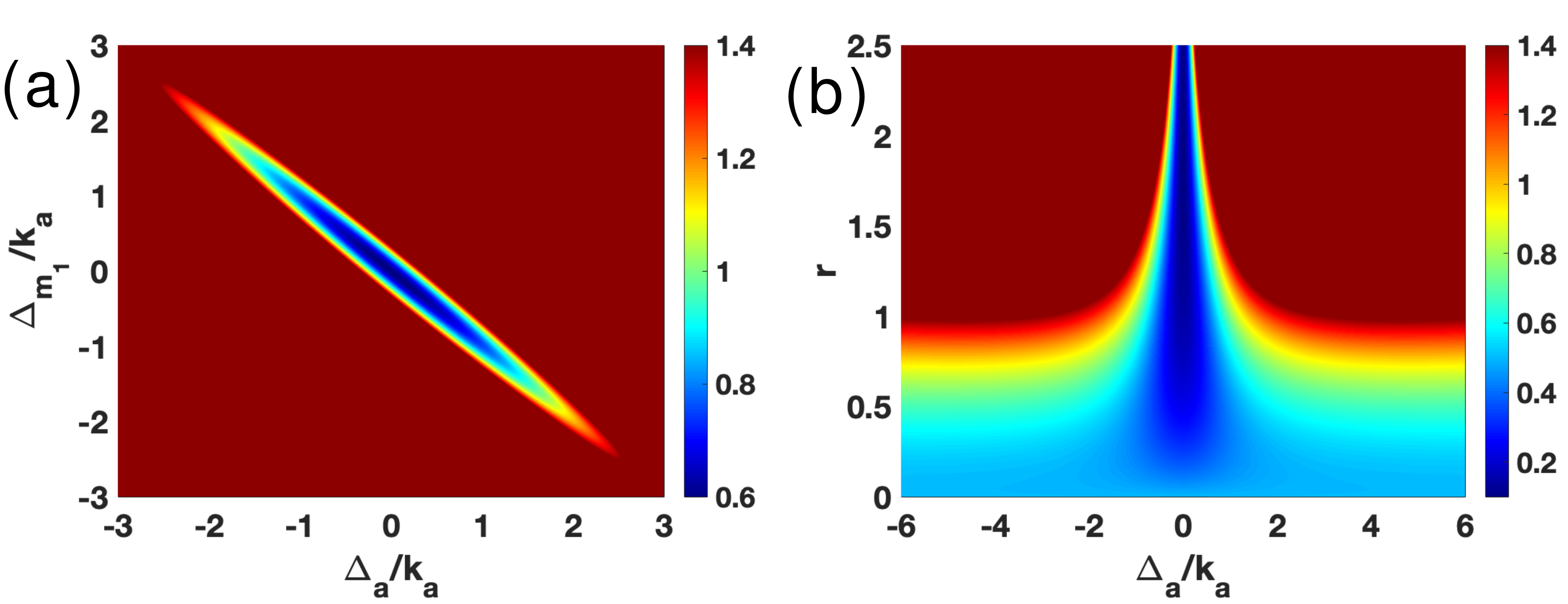}
\caption{\small  {(a) $\langle \delta M_{x}^{2}\rangle +\langle \delta m_{y}^{2} \rangle$ against $\Delta_{a}$ and $\Delta_{m_{1}}$ with $\Delta_{m_{2}}=\Delta_{m_{1}}$, $r=2$, $\theta=0$, $T=20$ mK. (b) $\langle \delta M_{x}^{2}\rangle$ against $\Delta_{a}$ and squeezing parameter $r$ with $\Delta_{m_{1}}=\Delta_{m_{2}}=0$, $\theta=0$, T=20 mK.}} \label{f3a}
\end{figure}

To compare our results with the protocols using nonlinear methods, a recent work \cite{n5} produced an entanglement close to 0.25 between the magnon modes at a temperature 10 mK through a Kerr nonlinearity introduced by a strong classical drive. The use of a different kind of nonlinearity, namely the magnetostrictive interaction in one YIG sphere produces similar entanglement \cite{n4} at a temperature 10 mK. The entanglement vanishes as the temperature approaches 20 mK. In contrast our scheme for entanglement generation produces a steady and strong entanglement between 0 to 100 mK and a significant amount of entanglement is present even at 500 mK. The mechanism of the entanglement generation will become clear from the discussion below.

Next we discuss two different criteria for entanglement in a two mode CV system. The advantage of these criteria over logarithmic negativity is that the former can be easily examined through experiments \cite{22,23}, though in a qualitative way. The first inseparability condition proposed by Simon \cite{43} and Duan \textit{et al.} \cite{42} is the sufficient condition for entanglement in a two mode CV system. We define a new set of operators $M=(m_{1}+m_{2})/\sqrt{2}$, $m=(m_{1}-m_{2})/\sqrt{2}$. The criterion suggests that if the two modes are separable then they should satisfy the following inequality
\begin{align}
\langle \delta M_{x}^{2}\rangle+\langle \delta m_{y}^{2} \rangle \geq 1, \label{8}
\end{align}
 where $\delta M_{x}$ and $\delta m_{y}$ are the fluctuations in the quadratures $M_{x}$ and $m_{y}$ defined as $M_{x}=(M+M^{\dagger})/\sqrt{2}$, $m_{y}=i(m^{\dagger}-m)/\sqrt{2}$. In other words, violation of the inequality in Eq.(\ref{8}) means the existance of entanglement between the two YIG samples. Figure \ref{f3a}(a) shows that  there is region around $\Delta_{a}=0$ and $\Delta_{m_{1}}=0$ (resonance) in which $\langle \delta M_{x}^{2}\rangle+\langle \delta m_{y}^{2} \rangle$ is less than one and it is a clear manifestation of the entanglement present between the YIG samples. Mancini \textit{et al.} \cite{44} derived another inequality which is useful in characterizing separable states. It suggests that if the two mode CV system is separable, then it should satisfy the following inequality
 \begin{align}
 \langle \delta M_{x}^{2}\rangle \langle \delta m_{y}^{2} \rangle \geq 1/4. \label{9}
 \end{align}
 Hence the violation of Eq.(\ref{9}) implies that the YIG samples are entangled. We use identical coupling strengths between the cavity and the two YIG samples. Therefore when $\Delta_{m_{1}}=\Delta_{m_{2}}=0$ the Hamiltonian of the system in the rotating frame of the drive can be written as 
 \begin{align}\
\mathcal{H}/\hbar&=\Delta_{a} a^{\dagger}a+\sqrt{2}g_{m_{1}a}(a+a^{\dagger})(M+M^{\dagger}). \label{10}
\end{align}
The Hamiltonian does not contain a term involving $m$ and $m^{\dagger}$.  Hence the fluctuations in $m$ will be equal to the fluctuations at time $t=0$. Since $m$ at $t=0$ is in the vacuum state (at low temperature 20 mK), we have $\langle \delta m _{y}^{2}\rangle=1/2$. Figure \ref{f3a}(b) shows that there is a region close to resonance where the quantity $\langle \delta M_{x}^{2}\rangle$ is less than $1/2$. This violates the inequality in Eq.(\ref{9}) and hence the two YIG samples are entangled. This further corroborates our results.
\begin{figure}[!t]
\captionsetup{justification=raggedright,singlelinecheck=false}
\centering
\includegraphics[scale=0.23]{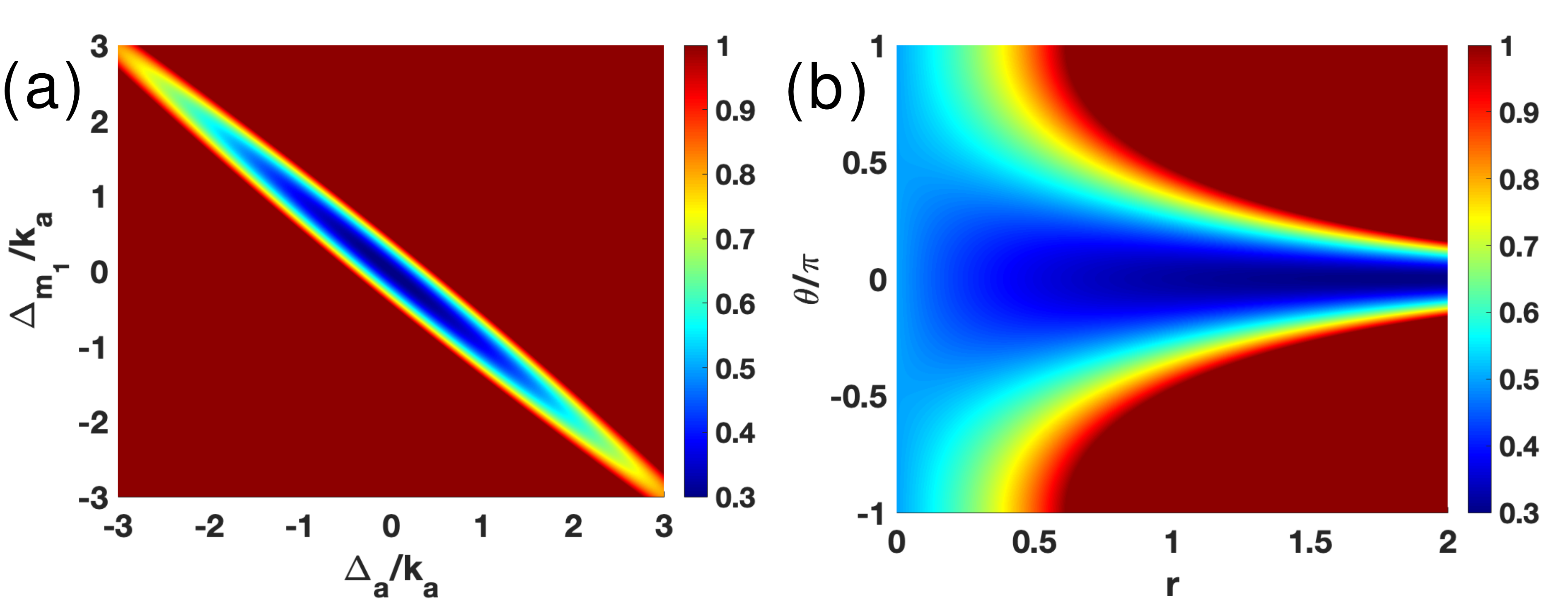}
\caption{\small  {(a) Variance of the first magnon quadrature $\langle \delta x_{1}(t)^{2}\rangle $ versus $\Delta_{a}$ and $\Delta_{m_{1}}$. (b) Variance of the first magnon quadrature against squeezing parameter $r$ and phase $\theta$. The other parameters in (a) are $r=2$, $\theta =0$, $\Delta_{m_{1}}=\Delta_{m_{2}}$, $T=$20 mK. Other parameters in (b) are $\Delta_{a}=\Delta_{m_{1}}=\Delta_{m_{2}}=0$ and $T=$20 mK. $\langle \delta x_{2}(t)^{2}\rangle $ is identical to $\langle \delta x_{1}(t)^{2}\rangle $.}}\label{f1a} 
\includegraphics[scale=0.23]{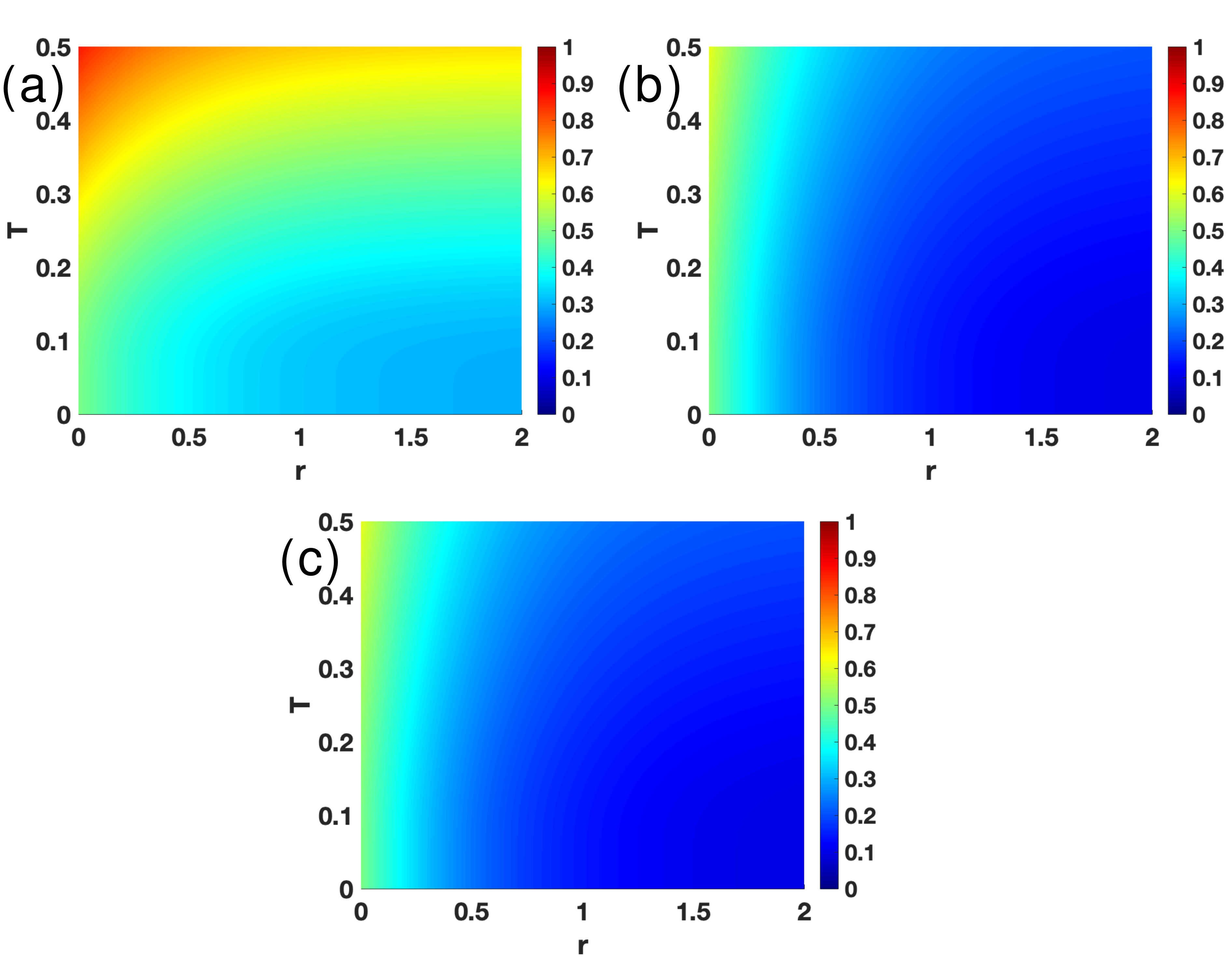}
\caption{\small  {(a) Variance of the first magnon quadrature $\langle \delta x_{1}(t)^{2}\rangle $ against squeezing parameter $r$ and temperature $T$ when both YIG samples are present. (b) Variance of the first magnon quadrature $\langle \delta x_{1}(t)^{2}\rangle $ against squeezing parameter $r$ and temperature $T$ with only one YIG sample is present. (c) Variance $\langle \delta M_{x}^{2}\rangle $ of the collective variable $M$ against squeezing parameter $r$ and temperature $T$. The other parameters are $\Delta_{a}=\Delta_{m_{1}}=\Delta_{m_{2}}=0$, $\theta=0$.}}\label{f1b}
\end{figure}
As a byproduct of our results we investigate the squeezing of the two magnon modes and show that it can be acheived by resonantly driving the cavity with a squeezed vacuum field. We are interested in the variances of the cavity and magnon mode quadratures and they are given by diagonal elements of the time-dependent CM $V$$(t)$ as defined previously. The amount of squeezing in a mode quadrature $X$ can be expressed in decibels (dB). It is obtained from the expression $-10\text{log}_{10}[\langle \delta X(t)^{2}\rangle/\langle \delta X(t)^{2}\rangle_{vac}]$, where $\langle \delta X(t)^{2}\rangle_{vac}=\frac{1}{2}$. As discussed in \cite{18} when the cavity and the two magnon modes are decoupled, the cavity field is squeezed as a result of the squeezed driving field and the magnon modes possesses vacuum fluctuations. As we increase the coupling strength, squeezing is partially transffered to the two identical YIG samples. The blue region in figure \ref{f1a}(a)-(b) represents the region of squeezing. For $r=2$ the input squeezing is about 17.35 dB. We observed a squeezing of about 2.27 dB for each of the two magnon modes at resonance with $T=$20 mK. Note that figure 5 give the magnon quadrature when both the YIG samples are present. Figure \ref{f1b}(a) shows that the magnon squeezing is robust against temperature. We observe moderate squeezing for both spheres even at $T=$0.35 K. At resonance we also find a squeezing of about 7.28 dB for the $M_{x}$ quadrature of the collective variable $M$. This is comparable to the results when one had only one YIG sample present and clearly manifested in figures \ref{f1b}(b)-(c). 

In conclusion, We have presented a scheme to generate an entangled pair of YIG samples in a cavity-magnon system. Entanglement of magnon modes can be generated through resonantly driving the cavity by a squeezed vacuum field and it can be realized using experimentally attainable parameters. The entanglement produced is robust against temperature. We observe considerable amount of entanglement even at $T$$=0.5$K. We have also discussed possible strategies to measure the generated entanglement. We have also showed that by employing the same method squeezed states of magnons in two different modes can be achieved. For an input squeezing of 17.35 dB we have observed a squeezing of about 2.27 dB for the magnon modes at $T$$=20$ mK.

Our scheme for entangling YIG samples does not require any nonlinearities and hence goes against the conventional wisdom of producing entanglement. This provides an entirely new method for entangling macroscopic systems, which can be used in other macroscopic systems.
\section{Acknowledgments}
Jayakrishnan would like to thank Jie Li for helpful discussions, carefully reading the article and providing constructive feedback.  


\begin{thebibliography}{8}
\bibitem{foot1}
However these nonlinearities are weak and require strong drives, thus the realization of entanglement is difficult. There have been exceptions especially in the context of superconducting qubits where strong Kerr nonlinearities have been used to produce a variety of cat states \cite{nn1}. Theoretical description for Kerr nonlinearity produced CAT states is given in \cite{nn2}.
\bibitem{foot}
We consider the size of the YIG sphere much smaller than the microwave wavelength and hence the effect of radiation pressure is negligible. The magnon-phonon coupling has been measured \cite{13} to be 8 or 9 orders smaller than the magnon-cavity coupling and hence we can neglect the effect of magnetostrictive forces.
\end{thebibliography}
\end{document}